%% file: paper.tex
\documentclass[sigconf, screen]{acmart}
\setcopyright{none}
\renewcommand\footnotetextcopyrightpermission[1]{}
\acmDOI{}
\RequirePackage{rotating}
\RequirePackage{fix-cm}
\usepackage{framed}
\usepackage[utf8]{inputenc}
\usepackage{subcaption}
\usepackage{multirow}
\usepackage{xspace}
\usepackage{ifthen}
\usepackage{url}
\usepackage{fancybox}
\usepackage{enumitem}
\usepackage{listings}
\usepackage[linesnumbered,boxruled]{algorithm2e}
\usepackage{algorithmic}
\usepackage{dirtytalk}
\usepackage{soul}
\usepackage{colortbl}
\usepackage{makecell}
\usepackage{siunitx}
\usepackage{fontawesome5}
\newcommand{\implication}[1]{\noindent\faHandPointRight[regular]\xspace\emph{#1}}

\newlist{rqlist}{description}{1}
\setlist[rqlist]{%
  font=\bfseries,
  style=nextline,
  labelsep=0.6em,
  leftmargin=!,
  widest={(RQ3)},
  itemsep=0pt,
  topsep=0pt,
  parsep=0pt,
  partopsep=0pt
}

\newcommand{\summaryblock}[2]{%
  \begin{oframed}
    \noindent\textbf{#1:} #2%
  \end{oframed}
}

\definecolor{mygray}{gray}{0.6}

\newlength\WIDTHOFBAR
\DeclareGraphicsExtensions{.pdf,.jpeg,.png}

	\newcommand{\eg}{e.g.,}
	\newcommand{\ie}{i.e.,}

\newcommand{\RQone}{How is the quality of taxonomies generated by automated approaches?}
\newcommand{\RQtwo}{How well do automatically generated taxonomies align with human-defined ones?}
\newcommand{\RQthree}{How reliably can generated taxonomies be applied by independent annotators?}
\newcommand{\RQfour}{How efficient are automated methods in terms of cost and runtime?}

\copyrightyear{}
\acmYear{2026}
\acmConference[ASE '26]{Proceedings of the 41st IEEE/ACM International Conference on Automated Software Engineering}{October 12--16, 2026}{Munich, Germany}
\acmBooktitle{Proceedings of the 41st IEEE/ACM International Conference on Automated Software Engineering (ASE '26), October 12--16, 2026, Munich, Germany}
\acmISBN{}

\begin{document}

\title[How Well Do LLMs Generate Taxonomies in the SE Domain? A Multi-perspective Evaluation Framework]{How Well Do LLMs Generate Taxonomies in the SE Domain?\newline A Multi-perspective Evaluation Framework}


\author{Sota Nakashima}
\affiliation{%
  \institution{Kyushu University}
  \city{Fukuoka}
  \country{Japan}
	}
\email{nakashima@posl.ait.kyushu-u.ac.jp}

\author{Yuta Ishimoto}
\affiliation{%
  \institution{The University of Osaka}
  \city{Osaka}
  \country{Japan}
	}
\email{ishimoto@ist.osaka-u.ac.jp}

\author{Masanari Kondo}
\affiliation{%
  \institution{Kyushu University}
  \city{Fukuoka}
  \country{Japan}
	}
\email{kondo@ait.kyushu-u.ac.jp}

\author{Tao Xiao}
\authornote{Corresponding author}
\affiliation{%
  \institution{Kyushu University}
  \city{Fukuoka}
  \country{Japan}
	}
\email{xiao@ait.kyushu-u.ac.jp}

\author{Yasutaka Kamei}
\affiliation{%
  \institution{Kyushu University}
  \city{Fukuoka}
  \country{Japan}
	}
\email{kamei@ait.kyushu-u.ac.jp}

%



\begin{abstract}
\input{section/00_eabst}
\end{abstract}
\vspace{-4mm}


\vspace{-4mm}
\maketitle

\input{section/01_intro}
\input{section/02_related}
\input{section/03_task_description}

\input{section/04_evaluation_framework}
\input{section/05_experimental_setup}
\input{section/06_result}
\input{section/07_discussion}
\input{section/08_threat}
\input{section/09_conclusion}
\vspace{-1mm}
\section*{Acknowledgments}
We gratefully acknowledge the financial support of: (1) JSPS for the KAKENHI grants (JP24K02921, JP25K03100, JP25K22845, JP26H02500, JP26K21198), (2) Japan Science and Technology Agency (JST) as part of Adopting Sustainable Partnerships for Innovative Research Ecosystem (ASPIRE), Grant Number JPMJAP2415, (3) the Kayamori Foundation of Informational Science Advancement for supporting Tao Xiao, and (4) the Inamori Research Institute for Science for supporting Yasutaka Kamei via the InaRIS Fellowship.
\input{section/data}

\bibliographystyle{ACM-Reference-Format}
\bibliography{references}

\end{document}

%% file: section/00_eabst.tex
Taxonomies provide a shared conceptual framework for organizing heterogeneous observations in software engineering (SE) research.
Manually constructing such taxonomies is labor-intensive and requires annotators with expertise in the SE domain.
While advances in Large Language Models (LLMs) have led to the emergence of automated taxonomy generation methods outside the SE domain, their applicability to technically complex SE artifacts remains unclear.
In this experience paper, we present the first comprehensive empirical evaluation of how state-of-the-art automated methods perform on SE artifacts through a multi-perspective evaluation framework, including taxonomy quality, alignment with taxonomies defined by human experts, reliability under independent annotation, and efficiency.
To support this evaluation, we systematically collect seven SE papers with publicly available artifacts and human-defined taxonomies, and conduct experiments using two automated methods (TnT-LLM and CLIMB) with five state-of-the-art LLMs.
Our evaluation reveals a clear trade-off:
TnT-LLM constructs high-quality taxonomies comparable to human-defined ones but incurs substantially higher cost and runtime and tends to generate overly complex taxonomies, whereas CLIMB is 15--40$\times$ faster and 8--49$\times$ cheaper but tends to score lower on quality when technical inference beyond surface-level similarity is required.
These findings suggest that TnT-LLM and CLIMB can be used in practical situations in the SE domain, while researchers should first assess the complexity of the generated taxonomies and their cost using a subset of the target data to decide whether to use automated methods or human experts.
Our work represents a first step toward a systematic understanding of automated taxonomy generation in SE, offering actionable insights for future research and practice.


%% file: section/01_intro.tex
\vspace{-2mm}
\section{Introduction}
\label{sec:intro}

A taxonomy is a hierarchical structure that organizes complex or heterogeneous information into coherent and meaningful categories. 
In software engineering (SE) research, taxonomies provide a framework for reporting and comparing empirical findings~\cite{usman2017taxonomies}.
While researchers can rely on existing taxonomies, emerging domains may lack established categorizations, requiring the construction of new ones to systematically understand the subject.

Despite their importance, taxonomy construction in SE faces three major challenges. 
First, it requires substantial manual effort, typically involving iterative coding and consensus building among multiple researchers~\cite{wang2025icse, shabani2025icse, kashiwa2022ist, ebrahimi2023blockchain}.
Such processes are costly and difficult to scale.
Second, it requires annotators with expertise in the SE domain, such as development processes, coding practices, and architecture.
Without sufficient expertise, taxonomies may include overlapping codes or vary across annotators, resulting in inadequate taxonomies~\cite{ebrahimi2023blockchain}.
Third, SE artifacts are arguably more complex than natural language artifacts: they comprise multiple formal and informal elements (\eg~source code, issue reports, build logs)~\cite{ahmed2025msr} that often need to be linked and interpreted together, and defining meaningful categories often requires technical inference beyond the explicit terms and descriptions contained in the artifacts.

Automating taxonomy generation using Large Language Models (LLMs) has the potential to address these challenges. 
By leveraging LLMs, human effort can be reduced.
Moreover, if LLMs possess sufficient domain knowledge and technical inference capabilities, they may generate taxonomies comparable to those created by SE experts. 
However, existing LLM-based automated methods have only been evaluated in domains outside SE~\cite{wan2024tnt, li2025climb}, and their applicability to technically complex SE artifacts remains unclear. 

Moreover, evaluating generated taxonomies is challenging because the perspectives to be prioritized vary with the intended purpose.
For example, reproducing prior SE studies requires correspondence with \textit{human-defined taxonomies}\footnote{We refer to \textit{human-defined taxonomies} as taxonomies constructed by the original paper authors, who have expertise in the target domain of each paper.} (alignment), and exploration under budget constraints makes generation cost critical (efficiency).
Since no single perspective captures these diverse goals, assessing automated taxonomy generation for SE artifacts requires a multi-perspective empirical evaluation.



In this paper, we construct a multi-perspective evaluation framework, including taxonomy quality, alignment with human-defined taxonomies, reliability under independent annotation, and efficiency. Based on this framework, we empirically investigate how far state-of-the-art automated methods in other domains can be applied to the SE domain.
To support this evaluation, we conduct a systematic literature review and curate a dataset comprising seven SE papers that provide eight human-defined taxonomies.
We then evaluate two automated approaches (TnT-LLM~\cite{wan2024tnt} and CLIMB~\cite{li2025climb}) with five state-of-the-art LLMs. 
This study reveals key challenges and limitations of taxonomy generation in the SE domain.


Specifically, we address the following research questions (RQs).

\begin{rqlist}
  \item[(RQ1) \RQone]
  \textbf{\textit{Motivation:}}
  High-quality taxonomies are essential for reliable coding and comparison in SE studies, yet it remains unclear whether automated methods developed outside SE can achieve comparable quality on SE artifacts.
  \\
  \textbf{\textit{Results:}}
  TnT-LLM achieves quality scores comparable to human-defined taxonomies, whereas CLIMB tends to score lower, especially on orthogonality-related criteria (e.g., the extent to which categories do not overlap in scope or content). 
  SE artifact characteristics influence quality: CLIMB underperforms when latent concept inference is required, while it yields comparable scores to human-defined taxonomies on datasets where categories align with surface-level cues in the artifacts.
  \item[(RQ2) \RQtwo]
  \textbf{\textit{Motivation:}}
  Taxonomies defined by SE researchers provide an established reference.
  Assessing alignment with them reveals whether automated methods recover comparable structures and category boundaries, increasing confidence in their practical use.
  \\
  \textbf{\textit{Results:}}
  Generated taxonomies show high category coverage against human-defined ones, and are stable across generator LLMs (HSR 0.874--0.898 for TnT-LLM, 0.770--0.854 for CLIMB). 
  TnT-LLM tends to recover more human-like categories, while CLIMB more often matches human-defined structure and scale.
  \item[(RQ3) \RQthree]
  \textbf{\textit{Motivation:}}
  Generated taxonomies should be usable in practice.
  Inter-annotator agreement reveals the practical limits of generated taxonomies, while its associations with taxonomy quality (RQ1) and alignment (RQ2) suggest what may improve reliability.
  \\
  \textbf{\textit{Results:}}
  Generated taxonomies are broadly usable (label utilization rate 0.859--0.981), while agreement varies across generator LLMs (0.567--0.647 for TnT-LLM, 0.489--0.586 for CLIMB). 
  Across datasets, agreement tends to decrease at deeper layers, and SE artifact characteristics also influence reliability. Higher agreement is associated with quality criteria (e.g., Non-overlap $\rho$=0.313, $p$<0.05) and structural alignment (CEDS $\rho$=0.461, $p$<0.001).
  \item[(RQ4) \RQfour]
  \textbf{\textit{Motivation:}}
  We evaluate efficiency to determine whether automated methods are feasible for real-world use in SE.
  \\
  \textbf{\textit{Results:}}
  Manual taxonomy construction takes hundreds of person-hours, whereas automated methods generate taxonomies in minutes to hours.
  CLIMB is more efficient than TnT-LLM across generator LLMs (8.173--49.441$\times$ cheaper and 15.403--40.443$\times$ faster). 
  Cost and runtime increase with the scale of the target taxonomy and the number of artifact instances, especially for TnT-LLM.
\end{rqlist}

In summary, this paper makes the following contributions:
\vspace{-1mm}
\begin{itemize}\setlength{\itemsep}{0pt}
  \item \textbf{Evaluation framework:} We present a multi-perspective framework for evaluating taxonomy generation on SE artifacts, jointly assessing quality, alignment with human-defined taxonomies, reliability, and efficiency.
  \item \textbf{Empirical study:} We conduct the first comprehensive empirical study of two state-of-the-art methods with five LLMs on eight SE datasets with human-defined taxonomies.
  \item \textbf{Actionable insights:}
  We derive insights into the applicability of automated methods in the SE domain and provide guidelines for their use in this domain.
\end{itemize}

%% file: section/02_related.tex
\vspace{-2mm}
\section{Related Work}
In this section, we position our study within the literature.

\vspace{-2mm}
\subsection{Taxonomy Construction in SE}
Taxonomies are constructed in SE research to capture and organize emerging phenomena and patterns~\cite{wang2025icse, shabani2025icse, kashiwa2022ist, ebrahimi2023blockchain}. 
Wang et al.~\cite{wang2025icse} constructed taxonomies to understand code generation errors made by LLMs, while Shabani et al.~\cite{shabani2025icse} proposed a taxonomy of flakiness causes to analyze Dockerfile flakiness.
Such taxonomies provide a shared conceptual framework for systematic analysis, comparison across studies, and cumulative knowledge building.

Despite their importance, manual taxonomy construction is labor-intensive and depends heavily on domain-specific knowledge and expertise of annotators.
Wang et al.~\cite{wang2025icse} spent 328 person-hours constructing two taxonomies from 687 instances,
whereas Kashiwa et al.~\cite{kashiwa2022ist} sampled 375 of 15,671 instances, highlighting the practical difficulty of examining an entire large corpus manually.
When labeling 50 instances, Ebrahimi et al.~\cite{ebrahimi2023blockchain} observed 12 inconsistencies and relied on the second author, who had greater expertise, to resolve them.
These limitations motivate reliable and scalable automated taxonomy construction in SE.

\input{figure/02_related/fig_background}

\vspace{-2mm}
\subsection{LLM-Assisted Taxonomy Construction in SE}
With advances in LLMs, an increasing number of studies~\cite{ahmed2025msr, yu2025autoempirical, abdeen2025taxonomic, nakashima2025apsec} have explored supporting or automating taxonomy construction in SE. 
We view this task as a two-step process: (1) \emph{taxonomy generation} and (2) \emph{category annotation}.
Figure~\ref{fig:background} illustrates the difference: in category annotation, a human-defined taxonomy is given and LLMs assign each SE artifact to one of its categories; in taxonomy generation, no pre-defined taxonomy exists and LLMs induce categories and their hierarchy directly from a collection of artifacts.

Prior work~\cite{ahmed2025msr, yu2025autoempirical} has mainly focused on automating \emph{category annotation}.
Ahmed et al.~\cite{ahmed2025msr} investigated whether LLMs can replace human annotation of SE artifacts.
In contrast, there is limited empirical evidence on how far automated taxonomy generation methods can produce taxonomies from scratch for SE artifacts.
Abdeen et al.~\cite{abdeen2025taxonomic} reported that taxonomies produced by a single LLM were insufficient and that obtaining a usable taxonomy required iterative, human-guided prompt refinement.
Our prior work~\cite{nakashima2025apsec} provided the first empirical investigation in SE on the extent to which LLMs can generate SATD taxonomies; however, the evaluation primarily focused on alignment with human-defined taxonomies.

Building on these findings, this study expands the empirical setting by curating a broader dataset and evaluating state-of-the-art methods originally proposed outside SE, revealing how their performance varies with the characteristics of SE artifacts.
Moreover, we introduce a multi-perspective framework for evaluating taxonomy generation from four complementary perspectives. 
Since the usefulness of a generated taxonomy depends on its intended purpose, no single metric can fully capture its value. 
Our framework supports goal-dependent evaluation, such as prioritizing alignment when comparing generated taxonomies with prior taxonomies.

\vspace{-2mm}
\subsection{Automated Taxonomy Generation}
To contextualize our work in SE, we review automated taxonomy generation methods developed \emph{outside} SE. 
Existing techniques are broadly categorized into two paradigms: top-down and bottom-up.

\noindent
\textbf{Top-Down Approaches.}
A significant body of non-SE research, spanning both the pre-LLM era~\cite{shen2018hiexpan, zhang2018taxogen, huang2020corel, shang2020nettaxo, lee2022taxocom} and the LLM era~\cite{zeng2024chain, gunn2024creating, marchenko2024taxorankconstruct, kargupta-etal-2025-taxoadapt}, follows a top-down strategy that expands a pre-existing seed taxonomy.
These approaches are effective for enriching established knowledge structures, but their dependence on a seed taxonomy makes them less suitable when building a taxonomy from scratch for a new domain.

\noindent
\textbf{Bottom-Up Approaches.}
Bottom-up methods aim to induce a taxonomy directly from a corpus, which aligns with our goal of constructing data-driven and adaptive taxonomies.
To the best of our knowledge, the state-of-the-art methods that operate in a fully automated, bottom-up, and seed-free manner are TnT-LLM~\cite{wan2024tnt} and CLIMB~\cite{li2025climb}. 
TnT-LLM is an LLM-powered end-to-end framework that first summarizes each input instance to obtain a compact representation and then iteratively generates, updates, and reviews categories to form a taxonomy. 
CLIMB first embeds and clusters the input instances to obtain coherent groups, and employs LLM-based agents to generate fine-grained categories for each cluster and to synthesize higher-level categories into a hierarchy.

Table~\ref{tab:domain_comparison} summarizes the differences between prior non-SE evaluations and our setting.
Prior non-SE evaluations focus on Natural Language (NL) inputs from a single source (\ie~conversation texts~\cite{wan2024tnt} and job postings~\cite{li2025climb}).
In contrast, SE artifacts require the simultaneous understanding of NL (e.g., issue reports) and Formal Language (FL) artifacts (e.g., source code).
Each instance is often collected from multiple sources (e.g., a Dockerfile with its build logs), which constrains automated methods to link and interpret these sources jointly.
Moreover, empirical SE analysis aims to characterize latent or previously unobserved phenomena, requiring technical inference beyond surface textual cues.
These differences make taxonomy generation in SE more challenging, and motivate our specialized empirical evaluation.

\input{table/02_related/domain_comparison}

%% file: figure/02_related/fig_background.tex
\begin{figure}[!t]
    \centering
    \includegraphics[width=0.93\linewidth]{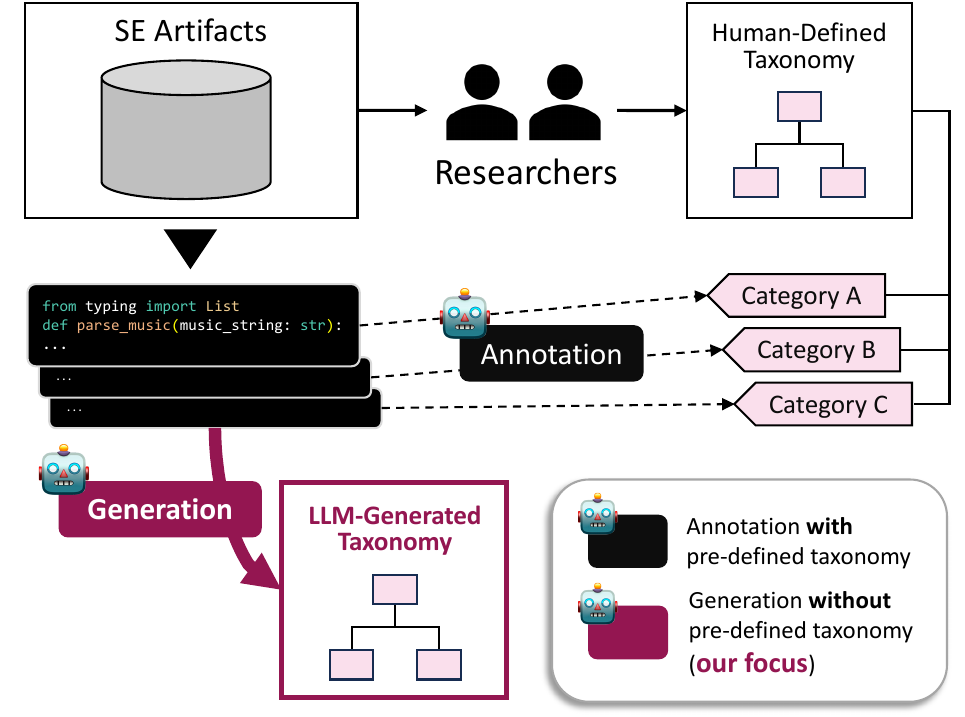}
    \vspace{-7pt}
    \caption{Taxonomy generation and annotation in SE}
    \label{fig:background}
    \vspace{-6pt}
\end{figure}

%% file: table/02_related/domain_comparison.tex
\begin{table}[t]
  \centering
  \caption{Comparison to related work}
  \vspace{-10pt}
  \label{tab:domain_comparison}
  \scriptsize
  \setlength{\tabcolsep}{4pt}

  \begin{tabular}{llll}
    \toprule
     & TnT-LLM~\cite{wan2024tnt} & CLIMB~\cite{li2025climb} & This Study \\
    \midrule
    \multirow{2}{*}{Use Case}
      & User Intent
      & \multirow{2}{*}{Labor Market}
      & \multirow{2}{*}{Empirical SE Analysis} \\
      & Conversational Domain
      & 
      & \\
    Input/Constraint
      & NL
      & NL
      & Mixed NL and FL \\
    Taxonomy Structure
      & One-Layer
      & Multi-Layer
      & Multi-Layer \\
    \bottomrule
  \end{tabular}
  \vspace{-10pt}
\end{table}

%% file: section/03_task_description.tex
\vspace{-2mm}
\section{Problem Definition}
\label{sec:task}

The input for taxonomy generation is a collection of SE artifacts, $A=\{a_1,a_2,\ldots,a_N\}$ and a task specification $S$.
Each artifact $a_i$ represents an SE artifact such as source code.
The task specification $S$ includes requirements such as the use case of the taxonomy (i.e., its intended purpose and analytical scope) and structural constraints (e.g., the number of categories and hierarchy layers).

Following prior work~\cite{zhang2018taxogen,zhang2025llmtaxo}, given $A$ and $S$, the goal is to build a tree-structured hierarchy $H$.
Each node $c \in H$ denotes a category that captures recurring concepts appearing in $A$ under the use case specified by $S$.
If $c$ has child categories $Ch(c)=\{c_1,c_2,\ldots,c_K\}$, each child $c_k \in Ch(c)$ should represent a sub-concept of $c$ and have a comparable semantic granularity to its siblings.

%% file: section/04_evaluation_framework.tex
\input{table/04_evaluation_framework/evaluation_criteria}

\vspace{-2mm}
\section{Evaluation Framework}
\label{evaluation_methods}
We construct an evaluation framework that integrates four complementary perspectives, providing a unified basis for assessing taxonomy generation methods:
(RQ1) \textbf{quality}, which captures the quality of the taxonomy itself;
(RQ2) \textbf{alignment}, which measures how closely generated taxonomies correspond to human-defined taxonomies;
(RQ3) \textbf{reliability}, which captures how consistently independent annotators can apply the taxonomy to the same corpus; and
(RQ4) \textbf{efficiency}, which captures the monetary cost and runtime of generating taxonomies with automated methods.

\noindent
\textbf{\textit{Quality:}}
Following prior work~\cite{zhang2025llmtaxo}, we measure quality by asking evaluator LLMs to score the taxonomy itself.
Each evaluator is given the taxonomy and the target use case, and rates 12 evaluation criteria on a 1--5 scale (Table~\ref{tab:intrinsic_quality}).
These criteria have four dimensions:
\textbf{Clarity}, which ensures that each category and its definition convey their meaning clearly so as to avoid confusion;
\textbf{Hierarchical Coherence}, which ensures that the taxonomy structure supports easy navigation and understanding by organizing categories from the most general to the most specific;
\textbf{Orthogonality}, which maintains clear boundaries between categories so that each captures a distinct aspect of the domain; and
\textbf{Completeness}, which captures how broadly and deeply the taxonomy covers the domain.
For single-layer taxonomies, some criteria are not applicable.
In such cases, we exclude Hierarchical Coherence dimension and Depth and Balance criteria under Completeness from evaluation.
The prompt for this quality evaluation is provided in our replication package~\cite{nakashima2026replication}.

\noindent
\textbf{\textit{Alignment:}}
Human-defined taxonomies were constructed by SE researchers with domain expertise and serve as a strong baseline; measuring alignment helps interpret whether the generated taxonomy recovers comparable concepts and structure.
This type of alignment evaluation is commonly adopted in prior work~\cite{shao2024assisting, franti2023soft_recall, zhu2023hierarchical, nakashima2025apsec}.
Following these studies, we use three complementary metrics: \textbf{Heading Soft Recall (HSR)}, \textbf{Catalogue Edit Distance Similarity (CEDS)}, and a normalized \textbf{Nodes Ratio (Nodes)}.

\textbf{Heading Soft Recall}~\cite{franti2023soft_recall} measures the proportion of categories in the human-defined taxonomy that are approximately covered by the generated taxonomy (on a scale from 0 to 1).
Rather than requiring exact string matches, it uses soft string similarity to allow minor lexical variations, such as paraphrases or different word forms.
Higher HSR values indicate that the generated taxonomy captures more conceptual categories in the human-defined taxonomy.
We focus on recall since our goal is to evaluate how well the generated taxonomy recovers these conceptual categories, rather than penalizing additional generated categories.

\textbf{Catalogue Edit Distance Similarity}~\cite{zhu2023hierarchical} evaluates the semantic and structural similarity between the generated taxonomy and the human-defined taxonomy by computing a normalized tree edit distance.
It measures both the hierarchical organization of taxonomies and the semantic similarity between their categories when computing the cost of matching nodes.
Higher CEDS values indicate greater semantic and structural similarity between two taxonomies.

To complement these metrics, we report the \textbf{Nodes Ratio}, defined as the number of nodes in the generated taxonomy divided by the number of nodes in the human-defined taxonomy.
This metric provides a coarse-grained indicator of structural scale differences between the two taxonomies.
Values close to 1 indicate that the generated taxonomy has a similar size to the human-defined one.

\noindent
\textbf{\textit{Reliability:}}
We measure reliability as the consistency with which independent annotators apply a taxonomy to the same corpus used for taxonomy generation.
Following prior work~\cite{li2025climb, ahmed2025msr, wan2024tnt, shah2023using}, we use a panel of LLMs as annotators because they enable scalable and cost-effective annotation.
For each generated taxonomy, we provide each LLM with the taxonomy and each SE artifact instance, and ask it to assign the leaf category that best corresponds to the artifact.
Each LLM independently follows the same procedure without access to the labels assigned by other annotators.
The prompts for this annotation are included in our replication package~\cite{nakashima2026replication}.

We quantify annotation consistency using \textbf{Hierarchical Agreement}, computed separately for each layer of the taxonomy.
For a given layer $k$, we compare the layer-$k$ labels assigned by annotators and treat them as agreeing when annotators select the same category at that layer.
This layer-wise computation captures cases where annotators agree at a coarse level (e.g., Layer~1) but disagree at a finer level (e.g., Layer~2).
Following prior work~\cite{ahmed2025msr, crupi2025effectiveness, haldar-hockenmaier-2025-rating}, we use Krippendorff's $\alpha$~\cite{krippendorff2018content} to measure agreement while accounting for chance; higher values indicate more reliable application.

Agreement alone may be high even when annotators repeatedly select a small subset of labels.
To complement agreement, we report \textbf{Label Utilization Rate}, defined as the percentage of taxonomy labels that are used at least once by annotators.
We compute utilization for each layer, analogous to our layer-wise agreement analysis.
A higher utilization rate suggests that annotators make use of the available distinctions rather than collapsing onto a few labels.

Prior work~\cite{wan2024tnt, shah2023using, li2025climb} also evaluates taxonomies using \textit{coverage}, typically defined as the percentage of instances assigned to a specific category rather than \textit{Other}.
However, this definition implicitly assumes that the input corpus is noise-free.
In our datasets, noisy instances exist and assigning an instance to \textit{Other} can be appropriate.
Therefore, we do not treat \textit{Other} as a special case; instead, we treat it as a regular category and do not report coverage.

\noindent
\textbf{\textit{Efficiency:}}
We assess efficiency using two metrics: \textbf{cost} and \textbf{runtime}.
For \textbf{cost}, we estimate the total API usage cost incurred by an automated method.
For \textbf{runtime}, we measure the end-to-end wall-clock time required to generate a taxonomy.


%% file: table/04_evaluation_framework/evaluation_criteria.tex
\begin{table*}[t]
  \centering
  \caption{RQ1 - Quality dimensions and evaluation criteria}
  \vspace{-10pt}
  \label{tab:intrinsic_quality}
  \scriptsize
  \setlength{\tabcolsep}{3pt}

  \begin{tabular}{p{2.5cm} p{3.0cm} p{2.2cm} p{6.4cm}}
    \toprule
    Dimension & Criterion & Abbreviation & Description \\
    \midrule
\multirow{4}{*}{Clarity}
  & Precision & CL-Pre & Categories use specific and well-defined terms. \\
  & Unambiguity & CL-Unamb & Each category has a single clear interpretation. \\
  & Consistency & CL-Cons & Terminology is used consistently across the taxonomy. \\
  & Accessibility & CL-Acc & Language is clear and avoids unnecessary jargon. \\
\midrule

\multirow{3}{*}{Hierarchical Coherence}
  & Gradational Specificity & HC-Grad & Categories progress logically from general to specific. \\
  & Parent--Child Coherence & HC-PC & Child categories logically belong to their parent. \\
  & Consistency & HC-Cons & Categories at the same level maintain comparable scope and specificity. \\
\midrule

\multirow{2}{*}{Orthogonality}
  & Distinctiveness & OR-Dist & Categories introduce meaningful distinctions. \\
  & Non-overlap & OR-NonOv & Minimal overlap in scope or content between categories. \\
\midrule

\multirow{3}{*}{Completeness}
  & Domain Coverage & CP-Cov & Coverage of a wide range of significant domain aspects. \\
  & Depth & CP-Depth & Sufficient depth within branches to capture nuanced distinctions. \\
  & Balance & CP-Bal & Even distribution of detail across branches. \\
    \bottomrule
  \end{tabular}
  \vspace{-8pt}
\end{table*}

%% file: section/05_experimental_setup.tex
\vspace{-2mm}
\section{Experimental Setup}

In this section, we introduce our experimental setup, including dataset collection, baselines, selected models, and prompt design.

\input{figure/05_experimental_setup/fig_dataset_collection}

\vspace{-2mm}
\subsection{Dataset Collection}
\label{data_collection}
To construct our dataset, we require (i) SE artifacts and (ii) a manually constructed taxonomy. 
The artifacts are used as input to automated approaches, while the human-defined taxonomy serves as a baseline. 
Following prior work~\cite{hou2024large, zhang2011identifying}, we conduct a systematic literature review to identify SE papers in which the researchers constructed such taxonomies (Figure~\ref{fig:data_collection}).
We collect papers from two major digital libraries: IEEE Xplore and ACM Digital Library. 
To mitigate potential data leakage for LLMs~\cite{balloccu-etal-2024-leak, dong-etal-2024-generalization}, we prioritize papers from as recent a period as possible (from January 1, 2024 to our search date, October 11, 2025).
Specifically, we use the following query to search for papers whose abstracts contained specific keywords: \textit{``taxonomy'' OR ``open coding'' OR ``manual coding'' OR ``qualitative''}. This query returned 9,456 papers.
After removing duplicates, we restrict the candidate set to papers published in the five top SE conferences and journals in Table~\ref{tab:venues}, resulting in 131 papers.
The first author screens the papers based on the full texts and replication packages using the following criteria:

\noindent\textbf{Inclusion criteria.}
\begin{itemize}\setlength{\itemsep}{0pt}
  \item The study targets a software engineering task (e.g., code generation, program repair and code review~\cite{hou2024large}).
  \item The study constructs a taxonomy through open coding (task, which inductively derives categories from data without pre-defined labels).
  \item The SE artifacts for reproducing the taxonomy construction are publicly available (e.g., replication packages).
\end{itemize}

\noindent\textbf{Exclusion criteria.}
\begin{itemize}\setlength{\itemsep}{0pt}
  \item The study assigns multiple categories to each input (e.g., ~\cite{tufano2024tse}).\footnote{We focus on single-label settings (one category per instance), as the taxonomy generation methods evaluated in this study are designed for that setting.}
  \item The required input context does not fit within the LLM context window (e.g., ~\cite{nourry2025my}).\footnote{Using the smallest context window (131k tokens for DeepSeek-V3.2) as the threshold, we exclude Nourry et al.~\cite{nourry2025my}, where 206 of 677 instances (30.4\%) exceeded the limit.}
  \item It is unclear which fields in artifacts to use as input (e.g.,~\cite{das2025tosem}).\footnote{Das et al.~\cite{das2025tosem} constructed multiple taxonomies. We exclude those with unclear input fields and retain only one as a dataset.}
\end{itemize}

\noindent
This screening yields seven papers with eight human-defined taxonomies in total; Table~\ref{tab:dataset} summarizes the dataset.

\input{table/05_experimental_setup/venue}

\input{table/05_experimental_setup/dataset}

\vspace{-2mm}
\subsection{Baselines}
\label{baselines}
In addition to the human-defined taxonomies collected in Section~\ref{data_collection}, we consider two state-of-the-art taxonomy generation methods proposed outside the SE domain as baselines.

\begin{itemize}
    \item \textbf{TnT-LLM}~\cite{wan2024tnt} is an LLM-powered end-to-end framework for automating text mining. It consists of (i) taxonomy generation and (ii) LLM-augmented text classification; we use only the taxonomy generation component. Since the original prompts target a flat taxonomy, we adapt them to generate hierarchical SE taxonomies via a recursive process that (1) generates top-level categories, (2) assigns documents to categories, and (3) repeats within each category.
    \item \textbf{CLIMB}~\cite{li2025climb} is a fully automated framework for generating occupation taxonomies from job postings by embedding and clustering instances, then using LLM-based agents to generate leaf categories and synthesize higher-level ones into a hierarchy. We apply this workflow to SE artifacts with prompts adapted to our SE use cases.
\end{itemize}

\textit{Implementation Details.}
Both TnT-LLM and CLIMB allow controlling the size of the generated taxonomy.
To enable a fair comparison, we instruct each method to match the structural properties of the corresponding human-defined taxonomy (i.e., the number of layers, top-level categories, and leaf categories).
For TnT-LLM, we split the instances into 20 mini-batches to accommodate the LLM context window within its multi-step generation pipeline.
For CLIMB, we apply the K-means algorithm to form clusters so that the number of clusters matches the specified number of leaf categories.

\vspace{-2mm}
\subsection{Selected Models}
\label{selected_models}
We use LLMs in two roles: \textbf{taxonomy generation} (as generators within TnT-LLM and CLIMB) and \textbf{evaluation} (as evaluators in RQ1 and annotators in RQ3).
To support these roles, we use closed and open models as LLMs.
Among closed models, we chose GPT-5.2 (\textit{gpt-5.2}) and Gemini-3-Pro (\textit{gemini-3-pro-preview}), which represent the latest flagship offerings from their respective providers.
From open models, we chose DeepSeek-V3.2 (\textit{deepseek-v3.2}), Llama 4 Maverick (\textit{llama-4-maverick}), and Mistral Large 3 (\textit{mistral-large-2512}), which are also the latest models from the open-model families.
We also use \textit{text-embedding-3-large} as the embedding model.
We access these models through the OpenRouter API~\cite{openrouter}, as its standardized interface simplifies the process of running identical experiments across multiple models.
For taxonomy generation, we use all five models, whereas for evaluation, we use the three best-performing models---GPT-5.2, Gemini-3-Pro, and DeepSeek-V3.2---as selected based on our preliminary study (Section~\ref{sec:preliminary_study}).

\input{figure/05_experimental_setup/fig_prompt_content}

\vspace{-2mm}
\subsection{Prompt Design}
\label{prompt_design}
As both automated approaches rely on LLMs, prompt design is critical for generating taxonomies tailored to SE artifacts.
As illustrated in Figure~\ref{fig:prompt_content}, our prompts consist of four key components.
For each dataset in Table~\ref{tab:dataset}, the input to TnT-LLM and CLIMB consists of the same SE artifact instances analyzed in the corresponding original SE study to construct its taxonomy.

\begin{itemize}\setlength{\itemsep}{0pt}
  \item \textbf{Data}: the SE artifacts provided to the model.
  \item \textbf{Data description}: a brief explanation of what each input instance represents and what information it contains.
  \item \textbf{Use case}: the intended purpose and scope of the taxonomy to be generated.
  \item \textbf{Taxonomy structure}: constraints on the taxonomy size, including the number of layers (depth), the number of top-level categories, and the number of leaf categories.
\end{itemize}
\noindent
The full prompts are provided in our replication package~\cite{nakashima2026replication}.

%% file: figure/05_experimental_setup/fig_dataset_collection.tex
\begin{figure}[!t]
  \centering
  \scriptsize
  \includegraphics[width=0.75\linewidth]{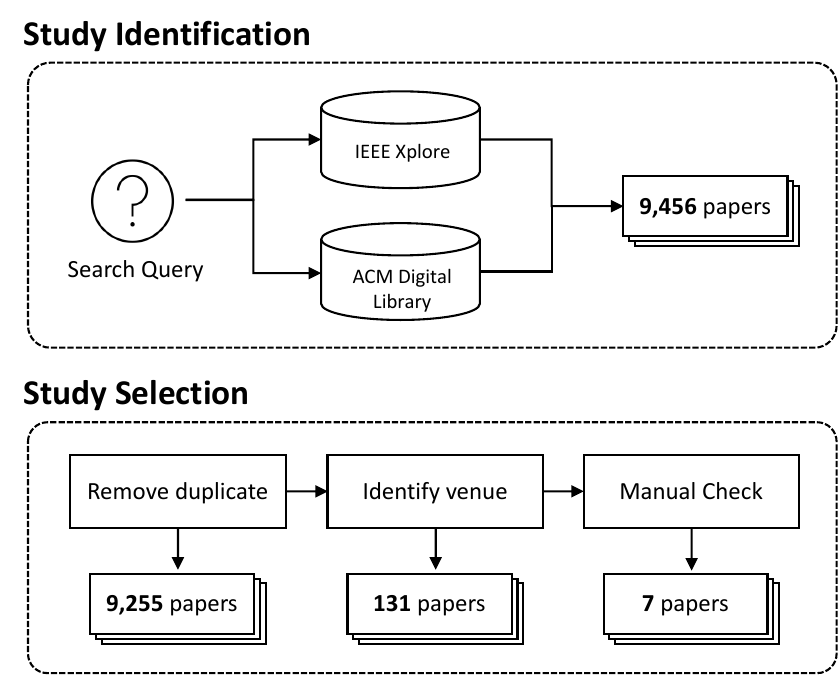}
  \vspace{-8pt}
  \caption{Study identification and selection process}
  \label{fig:data_collection}
  \vspace{-10pt}
\end{figure}

%% file: table/05_experimental_setup/venue.tex
\begin{table}[t]
\centering
\scriptsize
\caption{Publication venues for study selection}
\vspace{-10pt}
\label{tab:venues}
\begin{tabular}{ll}
\toprule
Acronym & Venues \\
\midrule
ICSE     & International Conference on Software Engineering \\
ASE      & International Conference on Automated Software Engineering \\
FSE      & International Conference on the Foundations of Software Engineering \\
TOSEM    & Transactions on Software Engineering and Methodology \\
TSE      & Transactions on Software Engineering \\
\bottomrule
\end{tabular}

\vspace{-5pt}
\end{table}

%% file: table/05_experimental_setup/dataset.tex
\begin{table}[t]
  \centering
  \caption{Datasets used in our study}
  \vspace{-10pt}
  \label{tab:dataset}
  \scriptsize
  \setlength{\tabcolsep}{3pt}
  \begin{tabular}{lcl}
    \toprule
    Dataset & \#Instance/\#Layer/\#Top/\#Leaf & Artifact \\
    \midrule
    Static Analysis Nondeterminism~\cite{miao2025icse}
      & 81/1/6/- & GitHub Issue \& commit \\

    LLM Code Error (Semantic)~\cite{wang2025icse}
      & 687/2/7/13 & Source code \\

    LLM Code Error (Syntactic)~\cite{wang2025icse}
      & 687/2/7/14 & Source code \\

    Dockerfile Flakiness~\cite{shabani2025icse}
      & 974/2/7/29 & Dockerfile \& build log \\

    Quantum Classical Issue~\cite{zappin2025icse}
      & 483/4/5/69 & Discussion thread \\

    Fairness API Issue Topic~\cite{das2025tosem}
      & 1{,}946/2/6/10 & GitHub Issue \\

    Common Sense-Violating Bug~\cite{fan2025tosem}
      & 982/2/3/23 & Bug report \\

    C++ OOP-Related Feature~\cite{wang2025tse}
      & 788/2/6/17 & Bug report \& code patch \\
    \bottomrule
  \end{tabular}
  \vspace{-10pt}
\end{table}

%% file: figure/05_experimental_setup/fig_prompt_content.tex
\begin{figure}[!t]
  \centering
  \includegraphics[width=0.95\linewidth]{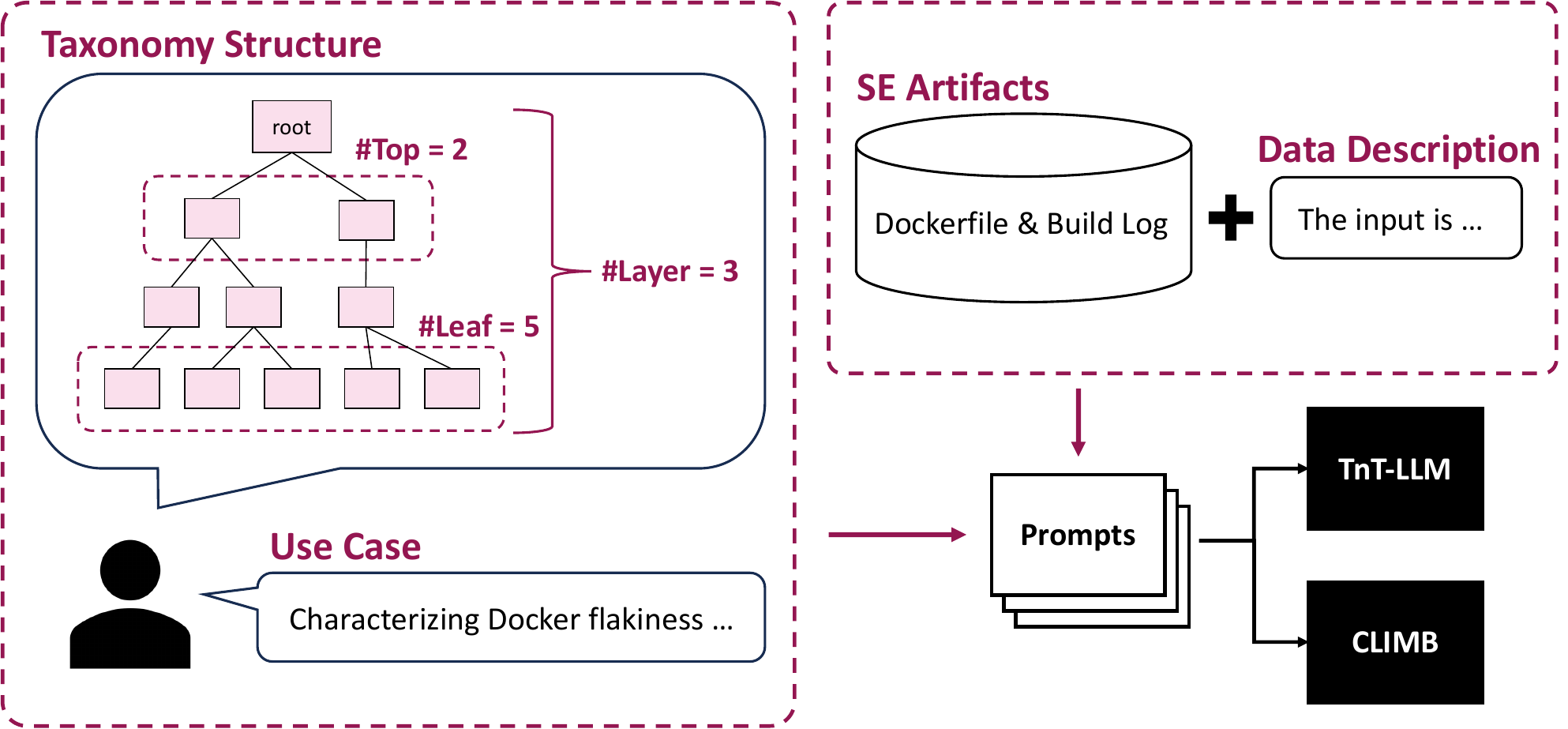}
  \vspace{-7pt}
  \caption{Example contents included in our prompts}
  \label{fig:prompt_content}
  \vspace{-5pt}
\end{figure}

%% file: section/06_result.tex
\input{section/06/06.0_preliminary_study}

\input{section/06/06.1_RQ1}
\input{section/06/06.2_RQ2}
\input{section/06/06.3_RQ3}
\input{section/06/06.4_RQ4}

%% file: section/06/06.0_preliminary_study.tex
\vspace{-2mm}
\section{Preliminary Study}
\label{sec:preliminary_study}
In RQ1, we use LLMs to score taxonomy quality.
To examine the reliability of this LLM-as-a-Judge setup~\cite{zheng2023judging, fu2024gptscore, liu2023g}, we manually validate it through (1) evaluator LLM selection on one dataset and (2) cross-dataset validation on the remaining seven datasets.

\vspace{-2mm}
\subsection{Evaluator LLM Selection}
\label{sec:preliminary_study_evaluator_selection}
To calibrate the evaluator LLMs, we manually analyze 11 taxonomies from \textit{LLM Code Error (Semantic)}: one human-defined taxonomy and ten generated taxonomies (two methods using five LLMs).
Two human evaluators and five evaluator LLMs rate each taxonomy against 12 quality criteria in Table~\ref{tab:intrinsic_quality} (11 taxonomies $\times$ 12 criteria = 132 ratings per evaluator).
Following prior work~\cite{ahmed2025msr}, if Human--Human and Human--LLM inter-rater agreement are comparable, evaluation outcomes are unlikely to differ systematically between human and LLM evaluators.
We therefore compare these agreements using Gwet's AC2~\cite{gwet2008computing} with quadratic weights, which is more robust than Cohen's $\kappa$ under imbalanced category distributions.

Table~\ref{tab:human_llm_agreement} shows that Human--LLM agreement for GPT-5.2, Gemini-3-Pro, and DeepSeek-V3.2 is comparable to Human--Human agreement (0.588), whereas Llama and Mistral exhibit substantially lower agreement, and we therefore excluded them from evaluator LLMs.

\input{table/06_results/preliminary/judge_reliability}

\vspace{-2mm}
\subsection{Cross-Dataset Validation}
\label{sec:preliminary_study_validation}
In Section~\ref{sec:preliminary_study_evaluator_selection}, our results show that GPT-5.2, Gemini-3-Pro, and DeepSeek-V3.2 achieve Human--LLM agreement comparable to Human--Human agreement on \textit{LLM Code Error (Semantic)}.
We examine whether this result generalizes to the remaining seven datasets in Table~\ref{tab:dataset}.
For each dataset, we include one human-defined taxonomy and the taxonomies generated by TnT-LLM and CLIMB using GPT-5.2---the best-performing generator overall across methods and quality criteria in RQ1---yielding 21 taxonomies in total.
Two human evaluators and the three previously selected evaluator LLMs independently assess each taxonomy (18 multi-layer taxonomies $\times$ 12 criteria $+$ 3 single-layer taxonomies $\times$ 7 applicable criteria = 237 ratings per evaluator).
For each dataset, we compute one Human--Human agreement and six Human--LLM agreements, one for each pairing of the two human evaluators with the three evaluator LLMs.
We then take the absolute difference between each Human--LLM agreement and the Human--Human agreement, and report the average of these six differences as the mean pairwise $|\Delta|$.

Table~\ref{tab:human_llm_agreement_validation} shows that the Human--LLM agreement is comparable to the Human--Human agreement across all seven datasets (mean pairwise $|\Delta| < 0.15$).
This result suggests that the selected evaluator LLMs remain aligned with human evaluations across the remaining datasets.
We therefore use these three models as evaluator LLMs in RQ1.
To maintain a consistent set of LLMs across our evaluation, we use the same three models as annotators in RQ3.

\input{table/06_results/rq1/judge_validation}

%% file: table/06_results/preliminary/judge_reliability.tex
\begin{table}[t]
    \centering
    \scriptsize
    \caption{Inter-rater agreement between two human evaluators and five evaluator LLMs on \textit{LLM Code Error (Semantic)}}
    \vspace{-10pt}
    \label{tab:human_llm_agreement}
    \begin{tabular}{lrrrrr}
    \toprule
     & GPT & Gemini & DeepSeek & Llama & Mistral \\
    \midrule
    Human Evaluator 1 & 0.529 & 0.711 & 0.672 & -0.146 & 0.189 \\
    Human Evaluator 2 & 0.519 & 0.677 & 0.541 & -0.149 & 0.117 \\
    \bottomrule
    \end{tabular}
    \vspace{-10pt}
    \end{table}


%% file: table/06_results/rq1/judge_validation.tex
\begin{table}[t]
    \centering
    \scriptsize
    \caption{Inter-rater agreement between two human evaluators and three selected evaluator LLMs across seven datasets}
    \vspace{-10pt}
    \label{tab:human_llm_agreement_validation}
    \setlength{\tabcolsep}{4pt}
    \begin{tabular}{lrr}
    \toprule
    Dataset & Human--Human & Mean Pairwise $|\Delta|$ \\
    \midrule
    Static Analysis Nondeterminism & 0.704 & 0.094 \\
    LLM Code Error (Syntactic) & 0.592 & 0.134 \\
    Dockerfile Flakiness & 0.665 & 0.142 \\
    Quantum Classical Issue & 0.458 & 0.137 \\
    Fairness API Issue Topic & 0.560 & 0.122 \\
    Common Sense-Violating Bug & 0.399 & 0.131 \\
    C++ OOP-Related Feature & 0.757 & 0.137 \\
    \bottomrule
    \end{tabular}
    \vspace{-10pt}
\end{table}

%% file: section/06/06.1_RQ1.tex
\input{figure/06_results/rq1/fig_quality_across_dataset}
\input{figure/06_results/rq1/fig_quality_across_model}

\vspace{-2mm}
\section{Quality of Generated Taxonomies (RQ1)}
In this section, we evaluate the quality of generated taxonomies.

\vspace{-2mm}
\subsection{Approach}
We assess taxonomy quality using 12 criteria spanning four dimensions: \textbf{clarity}, \textbf{hierarchical coherence}, \textbf{orthogonality}, and \textbf{completeness} (Table~\ref{tab:intrinsic_quality}).
As described in Section~\ref{sec:preliminary_study}, we use GPT-5.2, Gemini-3-Pro, and DeepSeek-V3.2 as evaluator LLMs and aggregate their scores for each criterion by taking the mean.
We analyze the results from two perspectives: (1) variation across datasets and (2) sensitivity to the choice of generator LLM.

\vspace{-2mm}
\subsection{Results}

\textbf{TnT-LLM and Human exhibit similar quality scores across datasets, while CLIMB tends to perform worse.}
Figure~\ref{fig:quality_across_dataset} shows that Human and TnT-LLM follow similar shapes across the twelve criteria, indicating comparable quality scores. In contrast, CLIMB frequently obtains lower values---particularly on orthogonality-related criteria (OR-Dist, OR-NonOv).
Compared to Human, TnT-LLM differs by only +0.018 on average across the twelve criteria, whereas CLIMB is lower by $-0.969$ on average.

\textbf{CLIMB is more likely to underperform when taxonomy generation requires inferring categories from SE artifacts.}
In \textit{Fairness API Issue Topic}, the baselines exhibit relatively similar scores, and both automated methods achieve scores close to the human baseline. 
This dataset is more topic-oriented, where categories can often be derived directly from surface-level cues in the artifacts. 
In such cases, artifacts with high semantic similarity are also likely to belong to the same category, allowing the embedding-based clustering of CLIMB to form coherent groups and achieve competitive quality scores. 
For other datasets, taxonomy generation requires inferring latent concepts from artifacts (e.g., flakiness causes from Dockerfiles and build logs). 
In these cases, artifacts belonging to the same conceptual category may not be similar at the surface level, making it harder to group them using artifact-level similarity. 
Consequently, generating categories from such clusters can lead CLIMB to produce overlapping categories (lower Non-overlap scores).

\textbf{The magnitude of quality differences between automated methods depends on the generator LLM.}
Figure~\ref{fig:quality_across_model} shows that the methods achieve relatively similar quality scores with GPT-5.2, whereas the gaps become larger with generators such as Llama 4 Maverick.
This is likely because taxonomy generation is open-ended: models propose categories themselves rather than choose from pre-defined labels.
Accordingly, capable generators keep quality scores across baselines relatively stable, whereas weaker ones expose method-dependent weaknesses more clearly.

\textbf{Generator sensitivity differs across quality dimensions, with label-level clarity appearing more robust than orthogonality.}
Figure~\ref{fig:quality_across_model} shows that the clarity-related criteria (e.g., CL-Pre) remain relatively stable across generators. 
In contrast, the orthogonality-related criteria (OR-Dist, OR-NonOv) show larger variation. 
This may reflect their greater sensitivity to generator-dependent decisions about splitting or merging concepts.

To enable readers to inspect and compare the exact scores underlying the trends shown in both RQ1 figures, we provide the detailed numerical results as tables in our replication package~\cite{nakashima2026replication}.

\summaryblock{Answer to RQ1}
{
TnT-LLM achieves quality scores comparable to Human across datasets, whereas CLIMB scores lower, especially on orthogonality-related criteria (e.g., Non-overlap). 
The gap varies by generator, being smaller for some (e.g., GPT-5.2) and larger for others (e.g., Llama 4 Maverick). 
SE artifact characteristics also matter: CLIMB is more likely to underperform when category generation requires inferring latent concepts (e.g., \textit{Dockerfile Flakiness}), whereas topic-oriented datasets (e.g., \textit{Fairness API Issue Topic}) show smaller gaps.
}

%% file: figure/06_results/rq1/fig_quality_across_dataset.tex
\begin{figure*}[!t]
    \centering
    \includegraphics[width=0.90\linewidth]{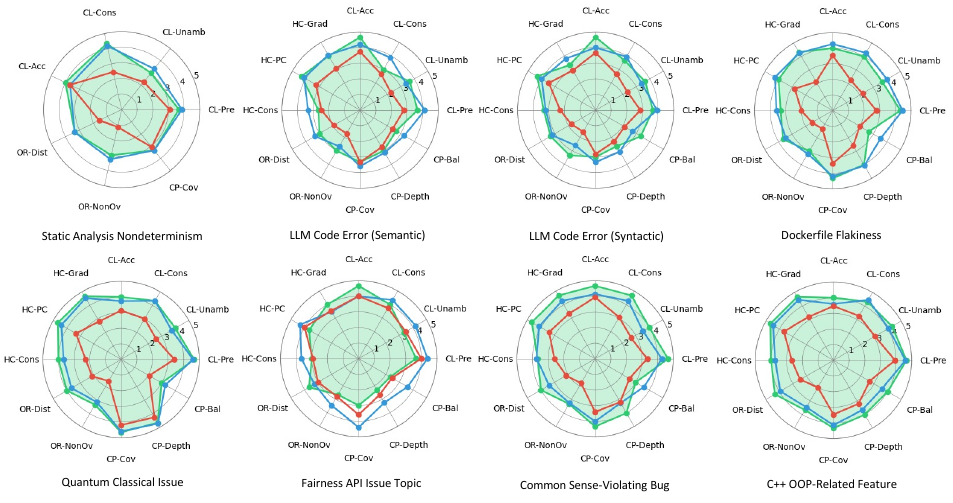}
    \vspace{-10pt}
    \caption{RQ1 - Average quality score aggregated by dataset}
    \label{fig:quality_across_dataset}
    \vspace{-5pt}
\end{figure*}

%% file: figure/06_results/rq1/fig_quality_across_model.tex
\begin{figure*}[!t]
    \centering
    \includegraphics[width=0.70\linewidth]{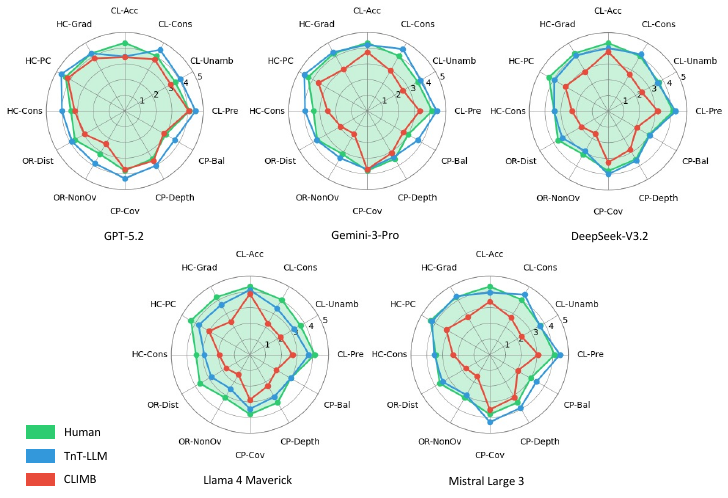}
    \vspace{-10pt}
    \caption{RQ1 - Average quality score aggregated by generator LLM}
    \label{fig:quality_across_model}
    \vspace{-5pt}
\end{figure*}

%% file: section/06/06.2_RQ2.tex
\section{Human-Defined Taxonomy Alignment (RQ2)}
In this section, we evaluate how well taxonomies generated by automated methods align with human-defined taxonomies.

\vspace{-2mm}
\subsection{Approach}
We measure alignment with human-defined taxonomies using three metrics: \textbf{Heading Soft Recall (HSR)}, \textbf{Catalogue Edit Distance Similarity (CEDS)}, and \textbf{Nodes Ratio (Nodes)}.
As in RQ1, we analyze the results from two perspectives: (1) variation across datasets and (2) sensitivity to the choice of generator LLM.

\vspace{-2mm}
\subsection{Results}

\textbf{Across datasets, TnT-LLM more often recovers human-like categories, whereas CLIMB more often matches human-like structure and size.}
Table~\ref{tab:alignment_dataset} shows that TnT-LLM achieves higher HSR, indicating that it more often recovers categories similar to human-defined ones.
In contrast, CLIMB often attains higher CEDS and Nodes Ratios closer to 1, suggesting that its taxonomies tend to be structurally closer to the human-defined ones.
This result is consistent with RQ1: TnT-LLM is competitive with Human on clarity-related criteria, which may help it recover categories closer to those in human-defined taxonomies.
Deep and structurally complex target taxonomies can further amplify these differences.
In \textit{Quantum Classical Issue}, TnT-LLM yields a Nodes Ratio of 3.341 and a CEDS of 18.171, indicating substantial over-generation and weaker structural alignment with the human-defined taxonomy.
CLIMB produces a taxonomy much closer to the human-defined one in both size (Nodes: 0.900) and structure (CEDS: 28.478).

\textbf{Alignment is not strongly affected by the choice of generator LLM.}
Table~\ref{tab:alignment_model} shows that the HSR values remain consistently high for both automated methods across generator LLMs.
For TnT-LLM, HSR ranges from 0.874 to 0.898, while CLIMB ranges from 0.770 to 0.854.
These narrow ranges suggest that alignment with human-defined taxonomies does not vary across generator LLMs.
A similar stability can also be observed for CEDS and Nodes Ratio.


\summaryblock{Answer to RQ2}
{
Generated taxonomies show high category coverage against human-defined ones, and are stable across generator LLMs (HSR 0.874--0.898 for TnT-LLM, 0.770--0.854 for CLIMB).
Across datasets, TnT-LLM tends to recover more human-like categories; CLIMB more often matches human-defined structure and scale.
}

\input{table/06_results/rq2/alignment_across_dataset}

\input{table/06_results/rq2/alignment_across_model}

%% file: table/06_results/rq2/alignment_across_dataset.tex
\begin{table}[t]
\centering
\scriptsize
\setlength{\tabcolsep}{2pt}
\caption{RQ2 - Alignment aggregated by dataset}
\vspace{-7pt}
\label{tab:alignment_dataset}
\begin{tabular}{lrrrrrr}
\toprule
 & \multicolumn{2}{c}{HSR}
 & \multicolumn{2}{c}{CEDS}
 & \multicolumn{2}{c}{Nodes} \\
\cmidrule(lr){2-3}\cmidrule(lr){4-5}\cmidrule(lr){6-7}
Dataset & TnT & CLIMB & TnT & CLIMB & TnT & CLIMB \\
\midrule
Static Analysis Nondeterminism  & \textbf{0.850} & 0.796 & 52.001 & \textbf{59.313} & 1.029 & \textbf{1.000} \\
LLM Code Error (Semantic)             & \textbf{0.870} & 0.846 & 35.261 & \textbf{39.265} & \textbf{0.933} & 0.876 \\
LLM Code Error (Syntactic)            & \textbf{0.863} & 0.846 & \textbf{29.714} & 29.110 & 1.209 & \textbf{0.936} \\
Dockerfile Flakiness                  & \textbf{0.976} & 0.806 & 30.781 & \textbf{33.046} & \textbf{0.930} & 0.822 \\
Quantum Classical Issue               & \textbf{0.962} & 0.786 & 18.171 & \textbf{28.478} & 3.341 & \textbf{0.900} \\
Fairness API Issue Topic              & \textbf{0.825} & 0.806 & 36.108 & \textbf{42.468} & 1.294 & \textbf{1.000} \\
Common Sense-Violating Bug            & \textbf{0.889} & 0.844 & 30.375 & \textbf{41.264} & 0.756 & \textbf{0.985} \\
C++ OOP-Related Feature               & \textbf{0.850} & 0.738 & \textbf{37.898} & 31.828 & 0.957 & \textbf{0.974} \\
\bottomrule
\end{tabular}
\vspace{-6pt}
\end{table}

%% file: table/06_results/rq2/alignment_across_model.tex
\begin{table}[t]
\centering
\scriptsize
\setlength{\tabcolsep}{4pt}
\caption{RQ2 - Alignment aggregated by generator LLM}
\vspace{-7pt}
\label{tab:alignment_model}
\begin{tabular}{lrrrrrr}
\toprule
 & \multicolumn{2}{c}{HSR}
 & \multicolumn{2}{c}{CEDS}
 & \multicolumn{2}{c}{Nodes} \\
\cmidrule(lr){2-3}\cmidrule(lr){4-5}\cmidrule(lr){6-7}
Model & TnT & CLIMB & TnT & CLIMB & TnT & CLIMB \\
\midrule
GPT-5.2            & 0.898 & 0.854 & 31.920 & 36.416 & 1.906 & 0.980 \\
Gemini-3-Pro       & 0.881 & 0.823 & 33.765 & 38.508 & 0.969 & 0.993 \\
DeepSeek-V3.2      & 0.894 & 0.813 & 31.429 & 37.477 & 1.372 & 0.888 \\
Llama 4 Maverick   & 0.874 & 0.782 & 34.500 & 42.332 & 1.039 & 0.859 \\
Mistral Large 3    & 0.882 & 0.770 & 37.330 & 35.751 & 1.244 & 0.963 \\
\bottomrule
\end{tabular}
\vspace{-6pt}
\end{table}

%% file: section/06/06.3_RQ3.tex
\input{table/06_results/rq3/reliability_across_dataset}

\vspace{-2mm}
\section{Reliability of Taxonomy Application (RQ3)}
In this section, we evaluate how reliably generated taxonomies can be applied by independent annotators.

\vspace{-2mm}
\subsection{Approach}
We assess taxonomy reliability using two metrics: \textbf{agreement} and \textbf{label utilization rate}.
We analyze the results from three perspectives: (1) variation across datasets, (2) sensitivity to the choice of generator LLM, and (3) relationships with taxonomy characteristics.
Specifically, we compute Spearman correlations~\cite{artusi2002spearman} between agreement and (i) quality criteria (RQ1) and (ii) alignment metrics (RQ2), and adjust $p$-values using the Benjamini--Hochberg procedure.

\vspace{-2mm}
\subsection{Results}
\textbf{Inter-annotator agreement decreases at deeper hierarchy layers, particularly for TnT-LLM.}
Table~\ref{tab:reliability_across_dataset} shows that, across datasets with multiple layers, agreement tends to decrease as the hierarchy becomes deeper.
A representative example is \textit{Quantum Classical Issue} (4 layers, 69 leaf categories), where agreement for TnT-LLM declines from 0.469 at Layer~1 to 0.372 at Layer~4, whereas CLIMB maintains relatively stable agreement across layers and remains higher at Layer~4 (0.469). 
This pattern is consistent with the findings of RQ2 on structural alignment: TnT-LLM can produce redundant or overly fine-grained distinctions at deeper layers, which may increase ambiguity for annotators, whereas CLIMB maintains more stable hierarchical structure.

\textbf{The characteristics of the SE artifacts themselves also influence reliability.}
As discussed in RQ1, artifacts that require more inference are harder to categorize consistently, and this challenge appears to carry over to reliability. 
Taxonomies in \textit{Fairness API Issue Topic} maintain comparatively high agreement across layers for all baselines, likely because category labels can often be matched directly to explicit terms in the issue text. 
In contrast, taxonomies in \textit{LLM Code Error} exhibit lower agreement overall, as latent error types must be inferred from source code generated by LLMs rather than identified from surface-level cues.

\textbf{Reliability varies across generator LLMs, while label utilization remains consistently high across models.}
Table~\ref{tab:reliability_across_models} shows that agreement exhibits variation across generator models (0.567--0.647 for TnT-LLM; 0.489--0.586 for CLIMB), suggesting that the choice of generator affects how consistently independent annotators select categories. 
In contrast, label utilization stays consistently high (0.859--0.934 for TnT-LLM; 0.947--0.981 for CLIMB), indicating that most generated labels are applied in practice.


\textbf{Reliability is associated with clear category boundaries and structural alignment to human-defined taxonomies.}
Table~\ref{tab:quality_reliability_corr} shows the correlations ($\rho$) between taxonomy property metrics and agreement at the deepest layer.
For quality (RQ1), four criteria (CL-Unamb, CL-Cons, OR-NonOv, and CP-Bal) show statistically significant, moderate positive correlations with agreement ($p<0.05$), suggesting that clearer categories, less overlapping categories, and more balanced taxonomies tend to yield higher agreement.
For alignment (RQ2), CEDS is significantly correlated with agreement ($\rho=0.461$, $p<0.001$), whereas HSR and Nodes are not, indicating that reliability is particularly related to structural correspondence with human-defined taxonomies.

\input{table/06_results/rq3/reliability_across_model}

\input{table/06_results/rq3/relationship_quality_and_reliability}

\summaryblock{Answer to RQ3}
{
Generated taxonomies are broadly usable (label utilization rate 0.859--0.981), while agreement varies across generator LLMs (0.567--0.647 for TnT-LLM, 0.489--0.586 for CLIMB).
Across datasets, agreement tends to decrease at deeper layers; SE artifact characteristics also influence reliability.
Higher agreement is associated with quality criteria (e.g., Non-overlap) and structural alignment (CEDS).
}

%% file: table/06_results/rq3/reliability_across_dataset.tex
\begin{table}[t]
\centering
\scriptsize
\setlength{\tabcolsep}{2pt}
\caption{RQ3 - Average agreement and label utilization aggregated by dataset}
\vspace{-7pt}
\label{tab:reliability_across_dataset}
\begin{tabular}{@{}llrrrrrr@{}}
\toprule
& &
\multicolumn{3}{c}{Agreement ($\alpha$)} &
\multicolumn{3}{c}{Label Utilization} \\
\cmidrule(lr){3-5}\cmidrule(lr){6-8}
Dataset & Layer & TnT & CLIMB & Human & TnT & CLIMB & Human \\
\midrule
Static Analysis Nondeterminism     & 1 &  0.835 &  0.702 &  0.777 &  0.841 &  0.886 &  0.905 \\
\midrule
LLM Code Error (Semantic)                & 1 &  0.490 &  0.372 &  0.557 &  0.908 &  0.933 &  0.952 \\
                                         & 2 &  0.435 &  0.348 &  0.567 &  0.940 &  0.917 &  0.949 \\
\midrule
LLM Code Error (Syntactic)               & 1 &  0.536 &  0.461 &  0.411 &  0.950 &  0.958 &  1.000 \\
                                         & 2 &  0.489 &  0.339 &  0.430 &  0.967 &  0.977 &  0.976 \\
\midrule
Dockerfile Flakiness                     & 1 &  0.778 &  0.719 &  0.761 &  0.892 &  0.994 &  0.857 \\
                                         & 2 &  0.698 &  0.548 &  0.563 &  0.956 &  0.975 &  0.856 \\
\midrule
Quantum Classical Issue                  & 1 &  0.469 &  0.650 &  0.534 &  0.844 &  0.831 &  1.000 \\
                                         & 2 &  0.464 &  0.664 &  0.488 &  0.932 &  0.838 &  0.865 \\
                                         & 3 &  0.425 &  0.676 &  0.446 &  0.875 &  0.848 &  0.885 \\
                                         & 4 &  0.372 &  0.469 &  0.429 &  0.793 &  0.935 &  0.889 \\
\midrule
Fairness API Issue Topic                 & 1 &  0.814 &  0.766 &  0.650 &  0.857 &  1.000 &  1.000 \\
                                         & 2 &  0.766 &  0.701 &  0.613 &  0.912 &  1.000 &  1.000 \\
\midrule
Common Sense-Violating Bug               & 1 &  0.673 &  0.723 &  0.817 &  0.787 &  1.000 &  1.000 \\
                                         & 2 &  0.627 &  0.649 &  0.640 &  0.940 &  0.991 &  1.000 \\
\midrule
C++ OOP-Related Feature                  & 1 &  0.649 &  0.572 &  0.729 &  0.941 &  1.000 &  1.000 \\
                                         & 2 &  0.605 &  0.469 &  0.663 &  0.974 &  0.996 &  1.000 \\
\bottomrule
\end{tabular}
\vspace{-6pt}
\end{table}

%% file: table/06_results/rq3/reliability_across_model.tex
\begin{table}[t]
\centering
\scriptsize
\setlength{\tabcolsep}{6pt}
\caption{RQ3 - Average agreement and label utilization rate at the deepest hierarchy layer aggregated by generator LLM}
\vspace{-7pt}
\label{tab:reliability_across_models}
\begin{tabular}{@{}lrrrr@{}}
\toprule
\multirow{2}{*}{Model}
& \multicolumn{2}{c}{Agreement ($\alpha$)}
& \multicolumn{2}{c}{Label Utilization} \\
\cmidrule(lr){2-3}\cmidrule(lr){4-5}
& TnT & CLIMB & TnT & CLIMB \\
\midrule
GPT-5.2              &  0.599 &  0.557 &  0.859 &  0.981 \\
Gemini-3-Pro         &  0.647 &  0.489 &  0.929 &  0.974 \\
DeepSeek-V3.2        &  0.567 &  0.498 &  0.920 &  0.948 \\
Llama 4 Maverick     &  0.621 &  0.586 &  0.934 &  0.947 \\
Mistral Large 3      &  0.583 &  0.524 &  0.934 &  0.948 \\
\midrule
Human                &        &  0.585 &        &  0.947 \\
\bottomrule
\end{tabular}
\vspace{-5pt}
\end{table}

%% file: table/06_results/rq3/relationship_quality_and_reliability.tex
    \begin{table}[t]
    \centering
    \scriptsize
    \caption{RQ3 - Spearman correlation between taxonomy properties and agreement}
    \vspace{-7pt}
    \label{tab:quality_reliability_corr}
    \renewcommand{\arraystretch}{1.15}
    \begin{tabular}{lrrlrr}
    \toprule
    Metric & $\rho$ & $p$ & Metric & $\rho$ & $p$ \\
    \midrule
    \quad CL-Pre   & 0.183  & 0.131          & \quad OR-NonOv & 0.313  & 0.030$^{*}$ \\
    \quad CL-Unamb & 0.275  & 0.030$^{*}$    & \quad CP-Cov   & $-$0.090 & 0.441 \\
    \quad CL-Cons  & 0.279  & 0.030$^{*}$    & \quad CP-Depth & $-$0.064 & 0.582 \\
    \quad CL-Acc   & 0.150  & 0.216          & \quad CP-Bal   & 0.292  & 0.030$^{*}$ \\
    \quad HC-Grad  & 0.147  & 0.242          & \quad HSR      & 0.189  & 0.140 \\
    \quad HC-PC    & 0.210  & 0.130          & \quad CEDS     & 0.461  & <0.001$^{***}$ \\
    \quad HC-Cons  & 0.254  & 0.062          & \quad Nodes    & $-$0.019 & 0.867 \\
    \quad OR-Dist  & 0.190  & 0.130          & & & \\
    \bottomrule
    \end{tabular}

    \vspace{0.5mm}
    \scriptsize{Adjusted $p$-value: $^{*}<0.05$, $^{**}<0.01$, $^{***}<0.001$.}
    \vspace{-5pt}
\end{table}

%% file: section/06/06.4_RQ4.tex
\vspace{-2mm}
\section{Efficiency of Automated Methods (RQ4)}
In this section, we evaluate the efficiency of automated approaches.

\vspace{-2mm}
\subsection{Approach}
We assess efficiency using two metrics: \textbf{cost} and \textbf{runtime}.
We estimate the API usage cost using the prices listed by the OpenRouter API~\cite{openrouter} as of October 2025.
For each API call, we count input and output tokens using the \texttt{tiktoken} Python library~\cite{tiktoken} and compute the cost using the corresponding pricing.
To obtain a conservative estimate, we do not apply any prompt-caching discounts and treat all requests as cache misses.
We analyze the results from three perspectives:
(1) comparison with manual taxonomy construction,
(2) robustness across generator LLMs, and
(3) variation across datasets.

\vspace{-2mm}
\subsection{Results}
\input{table/06_results/rq4/efficiency_across_model}

\textbf{Compared to manual taxonomy construction, automated methods can substantially reduce the human effort required to produce a taxonomy.}
Wang et al.~\cite{wang2025icse} reported that constructing two taxonomies of code generation errors made by LLMs required 328 person-hours.
In our setting, generating the corresponding two taxonomies takes 8,224.847 seconds ($\approx$2.285 hours) with TnT-LLM and about 330.419 seconds ($\approx$5.507 minutes) with CLIMB (Table~\ref{tab:runtime_cost_mean_datasets}).
Although this study focuses on taxonomy generation (not including category annotation), this contrast highlights the practical scalability benefits of automated taxonomy generation.

\textbf{CLIMB consistently outperforms TnT-LLM in both cost and runtime across generator LLMs.}
Table~\ref{tab:runtime_cost_mean_models} shows that although absolute values vary across models, the relative efficiency gap remains substantial: TnT-LLM is approximately 8.173--49.441$\times$ more expensive and 15.403--40.443$\times$ slower than CLIMB across models.
The cost gap is largest with GPT-5.2 (49.441$\times$), while the runtime gap is largest with DeepSeek-V3.2 (40.443$\times$).
This difference is expected because end-to-end overhead is dominated by LLM invocations.
TnT-LLM relies on iterative LLM calls throughout the pipeline, whereas CLIMB leverages embeddings and clustering to guide the process and requires fewer LLM calls overall.

\textbf{Cost and runtime depend on both the number of artifact instances and the scale of the target taxonomy, and this effect is more pronounced for TnT-LLM.}
Table~\ref{tab:runtime_cost_mean_datasets} indicates that datasets with more instances and deeper or wider taxonomies (i.e., more layers and leaf categories) require substantially more computation.
This trend is particularly clear for TnT-LLM, whose iterative generation strategy repeatedly invokes the LLM to refine categories at each layer.
\textit{Quantum Classical Issue} (483 instances, 4 layers, 69 leaf categories) incurs the highest cost and runtime for TnT-LLM (36.518\,USD and 6964.088\,seconds).
Although CLIMB is consistently more efficient than TnT-LLM, it also exhibits increased runtime on datasets with larger numbers of artifacts or deeper hierarchies.

\input{table/06_results/rq4/efficiency_across_dataset}

\summaryblock{Answer to RQ4}{
Manual taxonomy construction takes hundreds of person-hours, whereas automated methods generate taxonomies in minutes to hours.
CLIMB is more efficient than TnT-LLM across all generator LLMs (about 8.173--49.441$\times$ cheaper and 15.403--40.443$\times$ faster).
Cost and runtime depend on both the scale of the target taxonomy and the number of artifact instances, with a stronger increase for TnT-LLM.
}

%% file: table/06_results/rq4/efficiency_across_model.tex


\begin{table}[t]
\centering
\scriptsize
\setlength{\tabcolsep}{5pt}
\caption{RQ4 - Average cost and runtime aggregated by generator LLM}
\vspace{-7pt}
\label{tab:runtime_cost_mean_models}
\begin{tabular}{lrrrr}
\toprule
& \multicolumn{2}{c}{Cost [USD]}
& \multicolumn{2}{c}{Runtime [s]} \\
\cmidrule(lr){2-3}\cmidrule(lr){4-5}
Model & TnT-LLM & CLIMB & TnT-LLM & CLIMB \\
\midrule
GPT-5.2            & 27.341 & 0.553 & 4038.913 & 236.831 \\
Gemini-3-Pro       & 20.462 & 0.962 & 7462.706 & 484.506 \\
DeepSeek-V3.2      &  2.218 & 0.152 & 8498.950 & 210.146 \\
Llama 4 Maverick   &  1.087 & 0.133 & 1945.772 & 101.030 \\
Mistral Large 3    &  4.566 & 0.203 & 2515.615 & 106.705 \\
\bottomrule
\end{tabular}
\vspace{-8pt}
\end{table}

%% file: table/06_results/rq4/efficiency_across_dataset.tex
\begin{table}[t]
\centering
\scriptsize
\setlength{\tabcolsep}{5pt}
\caption{RQ4 - Average cost and runtime aggregated by dataset}
\vspace{-7pt}
\label{tab:runtime_cost_mean_datasets}
\begin{tabular}{lrrrr}
\toprule
& \multicolumn{2}{c}{Cost [USD]}
& \multicolumn{2}{c}{Runtime [s]} \\
\cmidrule(lr){2-3}\cmidrule(lr){4-5}
Dataset & TnT-LLM & CLIMB & TnT-LLM & CLIMB \\
\midrule
Static Analysis Nondeterminism  & 0.483  & 0.087 & 410.487  & 65.921  \\
LLM Code Error (Semantic)             & 4.563  & 0.165 & 4373.751 & 156.857 \\
LLM Code Error (Syntactic)            & 4.840  & 0.182 & 3851.096 & 173.562 \\
Dockerfile Flakiness                  & 9.715  & 0.421 & 5208.563 & 313.742 \\
Quantum Classical Issue               & 36.518 & 1.193 & 6964.088 & 504.865 \\
Fairness API Issue Topic              & 12.954 & 0.149 & 6951.349 & 139.119 \\
Common Sense-Violating Bug            & 6.702  & 0.213 & 5479.491 & 194.973 \\
C++ OOP-Related Feature               & 13.305 & 0.795 & 5900.305 & 273.710 \\
\bottomrule
\end{tabular}
\vspace{-10pt}
\end{table}

%% file: section/07_discussion.tex
\vspace{-2mm}
\section{Discussion}
\label{discussion}
In this section, we (i) analyze LLM output variability, (ii) present representative examples of generated categories, and (iii) summarize implications (see \faHandPointRight[regular] icon) for future research and practice.

\vspace{-2mm}
\subsection{LLM Output Variability}
\label{sec:variability_analysis}
We conduct a variability analysis to assess the stability of generation and evaluation outcomes since LLM outputs can vary across runs.

Repeating the full experiment across all eight datasets and five generator LLMs is prohibitively expensive (approximately 2,307\,USD just for five generations, calculated from Table~\ref{tab:runtime_cost_mean_datasets}).
We therefore select \textit{Static Analysis Nondeterminism}, the least expensive dataset.

Generation and evaluation introduce different sources of LLM variability, so we measure them separately.
For generation variability, we use RQ2 alignment metrics, which are deterministic given a generated taxonomy and the human-defined taxonomy.
We therefore repeat taxonomy generation five times and report the range of alignment scores (max--min across repetitions).
Quality scores in RQ1 depend on LLM-based evaluation. 
We therefore repeat the evaluator LLM scoring five times on each generated taxonomy and report the average score range across repetitions.
We do not repeat RQ3 because inter-annotator agreement is largely influenced by taxonomy clarity and structure, whose variability is captured by the repeated RQ1 evaluation and RQ2 generation.

Table~\ref{tab:variability_generation} shows that, within the same metric, CLIMB generally has smaller ranges than TnT-LLM, especially for structural metrics (CEDS and Nodes).
This result indicates that CLIMB produces more stable structural alignment across repeated runs.
This difference is expected because CLIMB uses embedding-based clustering and requires fewer LLM calls.
Evaluation variability of LLM-as-a-Judge in RQ1 is small on the 1--5 quality scale: the mean score range across repetitions is 0.263 for GPT-5.2, 0.272 for Gemini-3-Pro, and 0.503 for DeepSeek-V3.2.
We average scores across the three evaluator LLMs, which mitigates evaluator-specific variability in RQ1.

\input{table/07_discussion/variability_generation}
\input{figure/07_discussion/fig_case_study}

\vspace{-2mm}
\subsection{Representative Example}
RQ2 shows that CLIMB generates taxonomies closer to human-defined structure, whereas TnT-LLM tends to over-generate when producing structurally complex taxonomies, e.g., in \textit{Quantum Classical Issue}, the human-defined taxonomy has 92 nodes, while TnT-LLM generates on average 307.4 nodes and CLIMB 82.8 nodes.

Here, we examine generated categories to confirm two findings: (1) TnT-LLM generates human-like categories with quality comparable to Human, and (2) CLIMB produces redundant categories when latent inference is required.
We choose \textit{Dockerfile Flakiness}, which requires inferring flakiness causes from Dockerfiles and build logs, and focus on an \texttt{apt-get} update failure: ``\textit{Could not connect to archive.ubuntu.com:80, connection timed out.}''
Figure~\ref{fig:case_study} shows representative branches; the complete taxonomies are available in our replication package~\cite{nakashima2026replication}.
Human and TnT-LLM define similar categories---\textit{Timeout Issues} vs.\ \textit{Network Connectivity and Timeout Errors}.
All annotator LLMs select these categories---consistent with RQ1 (TnT-LLM achieves quality comparable to Human), RQ2 (human-like categories), and RQ3 (high agreement).
In contrast, CLIMB generates redundant categories such as \textit{Network Dependency Failures} and \textit{Remote Dependency Connection Flakiness}, leading annotators to split across categories---consistent with RQ1 (low orthogonality, e.g., Non-overlap) and RQ3 (low agreement).

\vspace{-2mm}
\subsection{Implication}
\textbf{Automated methods exhibit complementary strengths that motivate hybrid designs.}
Our results reveal a fundamental trade-off: the clustering-based method (CLIMB) improves efficiency and structural alignment but struggles to capture latent concepts, whereas the fully LLM-driven method (TnT-LLM) produces semantically richer categories at higher cost and over-generates in complex taxonomies.
Moreover, annotation reliability is associated with category clarity and structural alignment (RQ3).

\implication{For developers of taxonomy generation methods}, future pipelines should combine clustering for efficiency with targeted LLM reasoning for semantic abstraction and category naming.
They could also leverage annotation reliability, rather than category comparisons alone, as feedback for refining ambiguous or overlapping categories.

\textbf{Characteristics of SE artifacts and targeted taxonomy structures guide method selection and human post-processing.}
CLIMB performs competitively when categories follow explicit surface-level cues, but its quality decreases when category generation requires latent technical inference.
Furthermore, inter-annotator agreement drops at deeper layers when using both methods (RQ3).

\implication{For users of taxonomy generation methods}, use TnT-LLM for inference-intensive artifacts and CLIMB for artifacts with explicit cues when lower cost and runtime are prioritized.
Human reviewers should focus on deeper subcategories, and for CLIMB, check orthogonality to manually resolve overlapping categories.

\textbf{Automated methods should be evaluated for their intended purpose and generator LLM.}
No single perspective fully captures the usefulness of a generated taxonomy.
When a taxonomy is intended for subsequent annotation, \textbf{quality (RQ1)} and \textbf{reliability (RQ3)} should be prioritized to ensure interpretable and consistently applicable categories.
For reproducing prior SE taxonomies, \textbf{alignment (RQ2)} should be emphasized. 
When researchers need to explore candidate taxonomies quickly or operate under budget constraints, \textbf{efficiency (RQ4)} should be prioritized. 
Furthermore, because the performance also depends on the generator LLM, e.g., capable generators can yield stable quality scores across methods (RQ1), candidate methods should be evaluated with generators.

\implication{For developers and users of taxonomy generation methods}, our multi-perspective framework supports goal-dependent evaluation.
Using this framework, we recommend assessing methods with generator LLMs on a small subset of the target data to decide whether the generated taxonomy can be adopted directly, requires human post-processing, or should be constructed manually.

%% file: table/07_discussion/variability_generation.tex
\begin{table}[t]
\centering
\scriptsize
\setlength{\tabcolsep}{4pt}
\caption{Range (max $-$ min) across five generation runs}
\vspace{-7pt}
\label{tab:variability_generation}
\begin{tabular}{lrrrrrr}
\toprule
 & \multicolumn{2}{c}{HSR}
 & \multicolumn{2}{c}{CEDS}
 & \multicolumn{2}{c}{Nodes} \\
\cmidrule(lr){2-3}\cmidrule(lr){4-5}\cmidrule(lr){6-7}
Model & TnT & CLIMB & TnT & CLIMB & TnT & CLIMB \\
\midrule
GPT-5.2            & \textbf{0.032} & 0.042 & 11.170 & \textbf{2.853} & 0.143 & \textbf{0.000} \\
Gemini-3-Pro       & \textbf{0.079} & 0.105 & 12.528 & \textbf{2.385} & 0.286 & \textbf{0.000} \\
DeepSeek-V3.2      & 0.158 & \textbf{0.094} & 5.164 & \textbf{1.958} & 0.286 & \textbf{0.000} \\
Llama 4 Maverick   & 0.080 & \textbf{0.027} & 21.874 & \textbf{9.729} & 0.286 & \textbf{0.143} \\
Mistral Large 3    & 0.062 & \textbf{0.032} & 7.603 & \textbf{2.191} & 0.143 & \textbf{0.000} \\
\bottomrule
\end{tabular}
\vspace{-6pt}
\end{table}

%% file: figure/07_discussion/fig_case_study.tex
\begin{figure}[!t]
  \centering
  \includegraphics[width=0.85\linewidth]{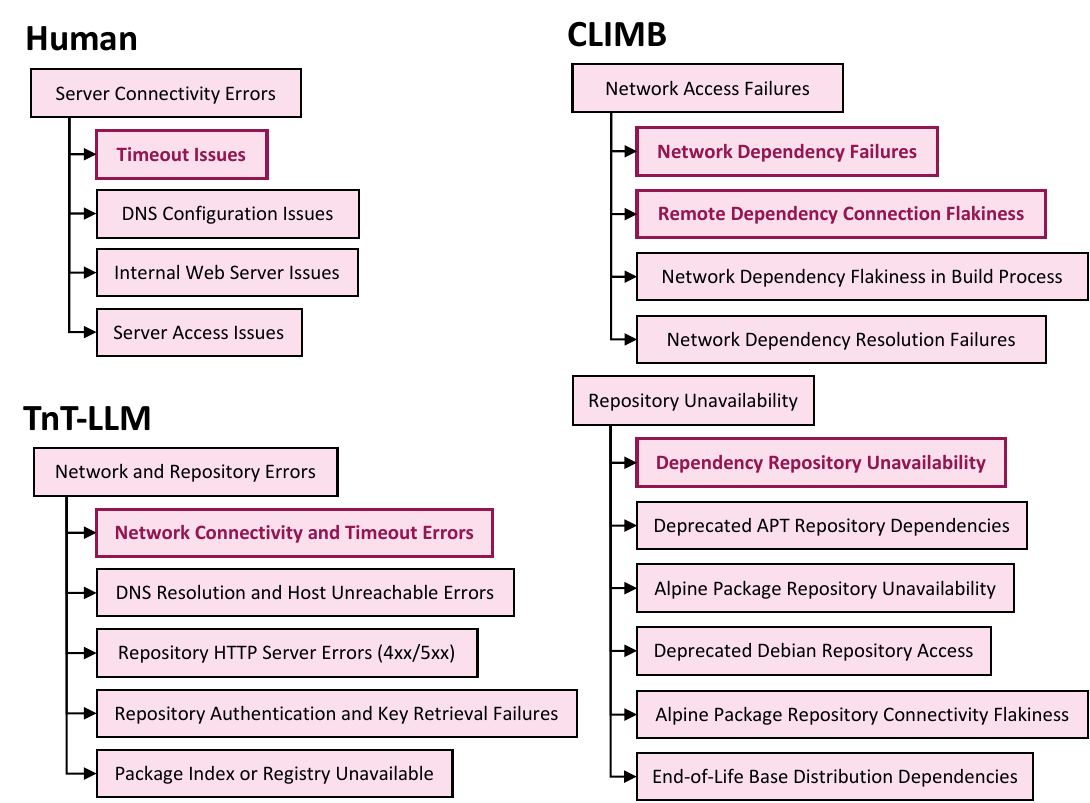}
  \vspace{-7pt}
  \caption{Representative branches of taxonomies for \textit{Dockerfile Flakiness}. Labels selected by annotators are highlighted.}
  \label{fig:case_study}
  \vspace{-12pt}
\end{figure}

%% file: section/08_threat.tex
\vspace{-2mm}
\section{Threats to Validity}
\label{validity}

\noindent
\textbf{Construct Validity.}
Our evaluation relies on LLMs to assess generated taxonomies at scale. 
While this design is cost-effective~\cite{zheng2023judging, fu2024gptscore, liu2023g}, LLM-based judgments may not perfectly reflect how human experts would assess taxonomy quality (RQ1) or how annotators would apply the same taxonomies in practice (RQ3). 


\noindent
\textbf{Internal Validity.}
Our LLM-based evaluation in RQ1 may be affected by \emph{self-bias}, whereby an LLM favors or penalizes taxonomies generated by the same model~\cite{crupi2025effectiveness, li2025preference, zheng2023judging, panickssery2024llm}.
Following prior work~\cite{crupi2025effectiveness}, for each evaluator and taxonomy, we computed the deviation from the mean score assigned by the other evaluators.
We then compared deviations for self-generated and other-generated taxonomies using Mann--Whitney U tests~\cite{mann1947test}.
We found no significant evidence of self-bias, suggesting limited impact on RQ1.

Our datasets were drawn from prior studies; some human-defined taxonomies may have appeared in the evaluated LLMs' training data.
To mitigate this risk, following prior work~\cite{li2024aaai}, we collected the most recent eligible studies available during data collection.

For fair comparison, we instruct automated methods to match the structure of human-defined taxonomies (e.g., \#Layer). 
However, different categories or hierarchies may still provide meaningful insights that were overlooked by human researchers.
Our findings may also depend on implementation choices (e.g., batching strategies).
In RQ3, we use zero-shot prompting; examining how prompting strategies~\cite{brown2020language, ahmed2022few, wei2022chain} affect the reliability of annotating SE artifacts with generated taxonomies is an important direction.

\noindent
\textbf{External Validity.}
Our findings may not generalize to other types of SE artifacts (e.g., requirements). 
Moreover, six of eight taxonomies have two layers, which may limit the generalizability of our findings.
Future work should establish a large-scale benchmark covering deeper, structurally diverse taxonomies and larger corpora.

%% file: section/09_conclusion.tex
\vspace{-2mm}
\section{Conclusion}
\label{sec:summary}
We present a multi-perspective evaluation framework for taxonomy generation in SE, jointly assessing quality, alignment, reliability, and efficiency.
Based on this framework, we reveal that TnT-LLM achieves quality comparable to human-defined taxonomies but is more costly, whereas CLIMB is 15--40$\times$ faster and 8--49$\times$ cheaper but performs worse when technical inference beyond surface-level similarity is required.
These results demonstrate the practical applicability of both methods while highlighting their distinct trade-offs.
We therefore recommend conducting a preliminary assessment using a subset of the target data before full-scale application, given their differing characteristics across four evaluation perspectives.


%% file: section/data.tex
\vspace{-1mm}
\section*{Data Availability Statement}
Our replication package is available online~\cite{nakashima2026replication}.